\documentclass[aps,pre,twocolumn,groupedaddress]{revtex4}
\usepackage[dvips]{graphicx}

\begin{document}
\title{Power-law friction in closely-packed granular materials}

\author{Takahiro Hatano}
\affiliation{Earthquake Research Institute, University of Tokyo, 1-1-1 Yayoi, Bunkyo, Tokyo 113-0032, Japan}

\date{\today}

\begin{abstract}
In order to understand the nature of friction in dense granular materials, 
a discrete element simulation on granular layers subjected to isobaric plain shear is performed.
It is found that the friction coefficient increases as the power of the shear rate, 
the exponent of which does not depend on the material constants.
Using a nondimensional parameter that is known as the inertial number, 
the power law can be cast in a generalized form so that the friction coefficients 
at different confining pressures collapse on the same curve.
We show that the volume fraction also obeys a power law. 
\end{abstract}

\maketitle

Friction is one of the oldest problems in science because it dominates various phenomena in our daily life.
In particular, dynamics of granular flow, which is ubiquitous in earth sciences and engineering, 
is governed by a law that describes behavior of the friction coefficient (ratio of the shear stress to the normal stress).
Such examples are avalanche, landslide, debris flow, silo flow, etc.
In addition, the nature of friction on faults, which plays a key role in earthquake mechanics \cite{aki,scholz}, 
is also attributed to that of granular rock because fault zone consists of layers of fine rock particles 
that are ground-up by the fault motion of the past.
To find a suitable law of friction in granular materials under a specific condition is thus an essential problem.

Although the frictional properties of granular materials are so important, our understanding is still limited.
In the context of earthquake mechanics, slip velocity (or shear rate) dependence of friction coefficient, 
which is equivalent to rheology under constant pressure condition, is a matter of focus \cite{scholz}.
In experiments on thin granular layers that are sheared at relatively low sliding velocities ranging from nm/s to mm/s, 
the behavior of the friction coefficient can be described by a phenomenological law 
in which friction coefficient logarithmically depends on the sliding velocity.
This is known as the rate and state dependent friction (RSF) law \cite{marone}.
Note that the RSF law also applies to friction at interfaces between two solids, as well as that in granular layers.
Although the RSF law applies well to lower speed (creep-like) friction, it is violated in high speed friction.
For example, several experiments indicated nonlogarithmic increase of the friction coefficient 
in granular layers at higher sliding velocities \cite{blanpied,kilgore,blanpied2}.
The same tendency was also observed in experiments on friction between two sheets of paper \cite{baumberger,heslot}.
However, at this point, we do not know any friction law that is valid at such higher velocities.

Several recent attempts to understand the nature of friction in granular media under high shear rates are noteworthy here.
Jop and the coworkers presented a simple friction law that describes flow on inclined planes \cite{jop}, 
based on massive simulations and experiments \cite{GDRmidi}. 
Although their friction law seems feasible, it involves rather dilute flow and 
its applicability to denser and slower flow (e.g. quasistatic flow) is not clear.
Da Cruz et al. \cite{dacruz} performed an extensive simulation that focused on dense and slow regime 
and found a friction law that does not contradict that of Jop et al.
However, because da Cruz et al. involved a two dimensional system, effect of the dimensionality may be questioned.
In particular, in quasistatic regime where the nature of interparticle contacts plays an esential role in rheology, 
the effect of dimensionality should be seriously investigated.

In this Rapid Communication, we perform a three dimensional simulation in order to understand 
the nature of friction in a slowly-sheared dense granular material.
Our particular interest is a dense granular matter under high confining pressure (e.g. tens of MPa)   
which is roughly corresponds to a typical configuration of faults at seismic slip.
Note that the RSF law is violated in such a situation.
A new law is reported in which friction coefficient increases as the power of shear rate.


In the following we describe the computational model of granular layers.
The individual constituents are assumed to be spheres, and their diameters range uniformly from $0.7d$ to $1.0d$.
The interaction force follows the discrete element method (DEM) \cite{cundall}.
Consider a grain $i$ of radius $R_i$ located at ${\bf r_i}$ with the translational velocity ${\bf v_i}$ 
and the angular velocity $\bf\Omega_i$.
This grain interacts with another grain $j$ whenever overlapped; i.e. $|{\bf r}_{ij}|<R_i+R_j$, 
where ${\bf r}_{ij}={\bf r}_i-{\bf r}_j$.
The interaction consists of two kinds of forces, each of which is normal and transverse to ${\bf r}_{ij}$, respectively.
Introducing the unit normal vector ${\bf n}_{ij}={\bf r}_{ij}/|{\bf r}_{ij}|$, 
the normal force acting on $i$, which is denoted by ${\bf F}^{(n)}_{ij}$, is given by 
$\left[f(\epsilon_{ij})+\zeta{\bf n}_{ij}\cdot\dot{{\bf r}}_{ij}\right]{\bf n}_{ij}$, where $\epsilon_{ij}=1-|{\bf r}_{ij}|/(R_i+R_j)$.
A function $f(\epsilon)$ describes elastic repulsion between grains.
Here we test two models: $f(\epsilon)=k \epsilon$ (the linear force) 
and $f(\epsilon)=k \epsilon^{3/2}$ (the Hertzian force) \cite{johnson}.
Note that the constant $k/d^2$ is on the order of the Young's modulus of the grains.
In order to define the transverse force, we utilize the relative tangential velocity ${\bf v}^{(t)}_{ij}$ 
defined by $(\dot{{\bf r}}_{ij}-{\bf n}_{ij}\cdot\dot{{\bf r}}_{ij})+(R_i{\bf\Omega}_i+R_j{\bf\Omega}_j)/(R_i+R_j)\times{\bf r}_{ij}$ 
and introduce the relative tangential displacement vector ${\bf \Delta}^{(t)}_{ij}=\int_{\rm roll} dt{\bf v}^{(t)}_{ij}$.
The subscript in the integral indicates that the integral is performed when the contact is {\it rolling}; 
i.e., $|\Delta^{t}_{ij}| < k_t$ or $\Delta^{t}_{ij}\cdot v^{t}_{ij} < 0$.
Then the tangential force acting on the particle $i$ is written as 
$-{\rm min}(\mu\dot{{\bf r}}_{ij}/|\dot{{\bf r}}_{ij}|, k_t{\bf \Delta}^{(t)}_{ij})|{\bf F}^{(n)}_{ij}|$.
In the case that $\mu=0$, the tangential force vanishes and the rotation of particles 
does not affect the translational motion.
The parameter values adopted in the present simulation are given in TABLE \ref{parameters}.
\begin{table}
\caption{The parameters of the discrete element simulation.}
\label{parameters}
\begin{center}
\begin{tabular}{ccccccc} \hline
polydispersity & $\zeta\sqrt{d/km}$ & $k_td$ & $\mu$ & $Pd^2/k$ \\ \hline
30 \% & 1 & 0.005 & 0 - 0.6 & $3.8\times10^{-5}$ - $1.1\times10^{-2}$ \\ \hline
\end{tabular}
\end{center}
\end{table}

The configuration of the system mimics a typical experiment on granular layers subjected to simple shear.
Note that there is no gravity in the system.
The system spans $L_x\times L_y \times L_z$ volume, and is periodic in the $x$ and the $y$ directions.
We prepare two systems of different aspect ratio, each of which contains approximately 10,000 particles: 
$25d\times25d\times 8d$, and $15d\times15d\times25d$.
As we shall discussed later, the difference of the aspect ratio does not affect the rheology.
In the $z$ direction, there exist two rough walls that consist of the same kind of particles as those in the bulk.
The particles that consist walls are randomly placed on the boundary and their relative positions are fixed.
The walls are parallel to each other and displaced antiparallel along the $y$ axis at constant velocities $\pm V/2$,
while they are prohibited to move along the $x$ axis.
One of the walls is allowed to move along the $z$ axis so that the pressure is kept constant at $P$.
Using the mass of the wall $M_w$ that is defined as the sum of the masses of the constituent particles, 
the $z$ coordinate of the wall $Z_w$ is described by the following equation of motion; 
$M_w\ddot{Z_w} = F_z - P S$, where ${\bf F}$ denotes the sum of the forces between 
the wall particles and the bulk particles, and $S$ denotes the area of the wall.
Then the $z$ component of the velocity of the wall particles is given by $\dot{Z}_w$.
Note that the friction coefficient of the system is defined by $F_y/PS$.

The system reaches a steady state after a certain amount of displacement of the walls.
We judge that the system reaches a steady state if each of the following quantities does not show apparent trends 
and seems to fluctuate around a certain value.
The monitored quantities are the friction coefficient, the $z$ coordinate of the wall (i.e., the density), 
and the granular temperature.
Also snapshots of the velocity profile are observed to ensure the realization of uniform shear flow.
We confirm that the transient behaviors of the friction coefficient and of the volume increase 
are quite similar to those observed in experiments.
Here we do not investigate such transients and restrict ourselves to steady-state friction.

Because uniform shear flow is unstable in a certain class of granular systems, 
we must check the internal velocity profiles at steady states.
There is a strict tendency that shear flow is localized near the walls in the case that 
the confining pressure is small and/or the sliding velocity is large.
This kind of spatial inhomogeneity is rather ubiquitous in granular flow, 
and is extensively investigated \cite{nott,saitoh}. 
In our simulation, uniform shear flow is realized at lower sliding velocities and higher confining pressures.
Here we discuss exclusively the case in which uniform shear is realized.
In this case, the shear rate is proportional to the sliding velocity of the walls; i.e., $\gamma=V/L_z$.

We investigate the behaviors of the friction coefficient of the system, $F_y / P S\equiv M$.
The control parameters that affect the friction coefficient are the shear rate $\gamma$ and the pressure $P$.
It is useful to represent the control parameters in terms of nondimensional numbers, 
because the friction coefficient is a nondimensional number and hence must be a function of nondimensional numbers.
Thus $\gamma$ and $P$ are recast in the following forms; $I=\gamma\sqrt{m/Pd}$ and $\Pi=Pd^2/k$.
In particular, the former is referred to as the inertial number \cite{ancey}, which dominates the frictional behavior of granular flows.
Hereafter we discuss the nature of friction taking advantage of these nondimensional numbers.

\begin{figure}
\includegraphics[scale=0.42]{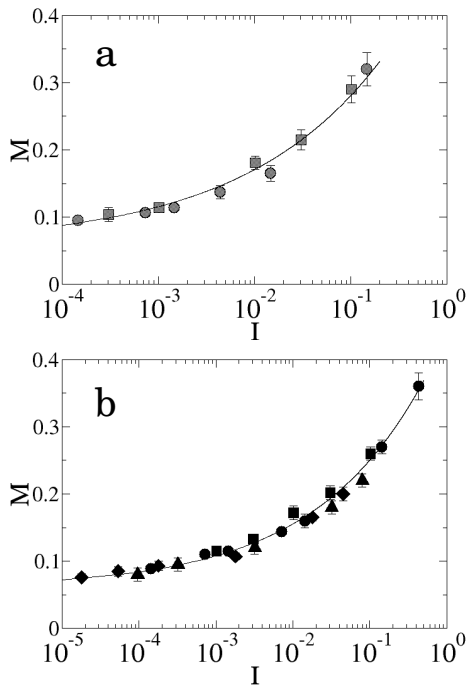}
\caption{
The friction coefficients $M$ in the models without tangential force, i.e., $\mu=0$.
The horizontal axis $I$ denotes the nondimensional shear rates, $V/L\sqrt{m/Pd}$.
The shape of the symbols and the confining pressure are in one-to-one correspondence: 
the squares to $\Pi=3.8\times10^{-5}$, the circles to $\Pi=1.9\times10^{-3}$, the triangles to $\Pi=3.9\times10^{-3}$, 
and the diamonds to $\Pi = 1.1\times10^{-2}$.
The layer thickness $L_z/d\simeq 8$ for the squares and the circles, 
while $L_z/d\simeq 26$ for the triangles and the diamonds.
The solid lines denote Eq. (\ref{powerlaw}) with $\phi=0.3$.
(a) Friction coefficient of the Hertzian force model.
(b) Friction coefficient of the linear force model.}
\label{mu0}
\end{figure}

In order to grasp the main point of our result, it is convenient to begin with the frictionless particles; i.e., $\mu=0$.
Recall that we test two models, each of which has different interaction: the Hertzian contact model and the linear force model.
As shown in FIG. \ref{mu0}, friction coefficients of these two models are collapsed on the following master curve.
\begin{equation}
\label{powerlaw}
M = M_0+s I^{\phi}, 
\end{equation}
where $M_0$ denotes the friction coefficient for $\gamma\rightarrow 0$. Here $M_0\simeq 0.06$.
Note that the effect of the inertial number is expressed by a power law, $s I^{\phi}$, where $\phi = 0.27\pm 0.05$.
The prefactor $s$ is approximately $0.37$ in the linear force model, 
while it is somewhat larger ($s\simeq0.45$) in the Hertzian contact model.

\begin{figure}
\includegraphics[scale=0.25]{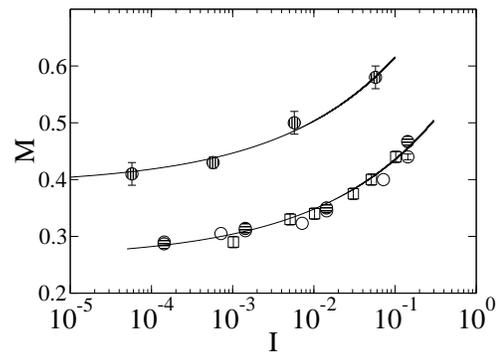}
\caption{
The friction coefficients in the models with tangential force ($\mu=0.2$, $0.6$).
The shape of the symbols and the confining pressure are in one-to-one correspondence as in FIG. \ref{mu0}.
The blank symbols denote the friction coefficient of the linear force model with $\mu=0.2$, 
while the symbols of vertically-striped pattern denote that of $\mu=0.6$.
The circles of horizontally-striped pattern denote the Hertzian force model with $\mu=0.2$.
The lines denote Eq. (\ref{powerlaw}) with $\phi=0.3$.}
\label{VM}
\end{figure}

In order to check the universality of Eq. (\ref{powerlaw}), we wish to confirm independence of our results on the details of the model.
First we discuss the effect of the tangential force between particles.
In FIG. \ref{VM}, shown are the friction coefficients of the models in which $\mu=0.2$ and $0.6$.
It is noteworthy that they range from $0.3$ to $0.4$, which are not significantly discrepant from those obtained by 
an experiment on spherical glass beads \cite{anthony}.
More importantly, the friction coefficients again obey Eq. (\ref{powerlaw}) with $\phi\simeq 0.3$ 
regardless of the value of $\mu$ and the force model (the linear or the Hertzian).
Indeed the friction coefficients of the both models are almost the same.
We also remark that the factor $s$ does not depend on $\mu$; $s=0.33\pm 0.03$ for $\mu=0$, $0.2$, and $0.6$.

On the other hand, $M_0$ depends on $\mu$.
In the linear force model, $M_0\simeq 0.06$ for $\mu=0$, while $M_0\simeq 0.26$ for $\mu=0.2$, and $M_0\simeq 0.4$ for $\mu=0.6$.
Similar dependence was also observed in Refs. \cite{campbell,dacruz}.
Note that $M_0$ also depends on the force model in the case that $\mu=0$; 
$M_0\simeq 0$ in the Hertzian model while $M_0$ is not negligible in the linear model.
Although $M_0$ looks like the static friction coefficient, note that it is defined in the $\gamma\rightarrow0$ limit 
and different from the static friction coefficient, above which a static system begins to flow.
In order to distinguish the two concepts, $M_0$ is referred to as dynamic yield strength.
The difference is important when we consider the stability of slip, as will be discussed in the last paragraph of 
this Rapid Communication.

It is important to notice that the data at different confining pressures collapse on the same curve by virtue of the inertial number.
This suggests that the friction coefficient of a dense granular material does not depend on $\Pi$, 
as long as it is small (in the present simulation $3.8\times10^{-5} \le \Pi \le 1.1\times 10^{-2}$).
Indeed, da Cruz et al. \cite{dacruz} found that the friction coefficient is independent of $\Pi$ 
($\kappa^{-1}$ in their notation) up to $\Pi \le 2.5\times 10^{-2}$ in a two dimensional system.
Therefore the independence on $\Pi$ is very likely within the accuracy of these simulations.
While one can still expect that the dependence may appear for larger $\Pi$, 
such a case that $\Pi\sim 0.1$ is meaningless as a model of a granular material.
Note that $\Pi$ roughly corresponds to the average overlap length of contacts divided by the particle diameter; 
namely. the average strain of individual particles.

Then we discuss the effect of inelasticity that is modeled by the viscous coefficient $\zeta$.
The corresponding nondimensional number $\tilde{\zeta}$ is defined by $\zeta\sqrt{d/km}$.
We find that decrease of $\tilde{\zeta}$ reduces the friction coefficient in the region where $I \gtrsim 0.01$, 
while the frictional strength is independent of $\tilde{\zeta}$ for smaller $I$ region.
This behavior is consistent with those obtained in Refs. \cite{campbell,dacruz}.
Nevertheless, it can be still described by Eq. (\ref{powerlaw}) with $s$ being a smaller value.
For example, the friction coefficient of a system in which $\tilde{\zeta}=0.05$ is described by 
$s\simeq 0.27$ with almost the same values of $M_0$ and $\delta$.
However, the functional form of $s(\tilde{\zeta})$ is not clear at this point.

From the discussions so far, we can conclude that the details of the present model 
do not affect the validity of Eq. (\ref{powerlaw}), which is the main result of this study.
Importantly, the exponent $\phi$ seems to be universal; it is approximately $0.3$ 
regardless of the details of the model and the control parameters.
The velocity-strengthening nature of this friction law does not contradict experiments 
\cite{blanpied,kilgore,blanpied2,baumberger,heslot}.
In addition, it illustrates universality of power-law in rheological properties of random media 
\cite{sollich} including foams \cite{gopal} and human neutrophils \cite{tsai}.
In the following we discuss four important points that are peripherally related to the main result.

\begin{figure}
\includegraphics[scale=0.25]{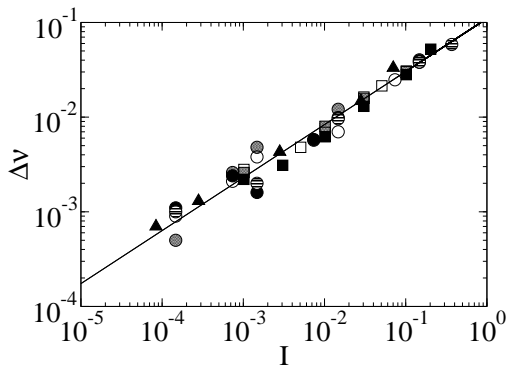}
\caption{The dilatation law. Decrease of the volume fraction $\Delta\nu =\nu_0 - \nu$ is plotted as a function of the inertial number.
Note that $\nu_0$ is the volume fraction in the $\gamma\rightarrow 0$ limit, which is estimated by extrapolation. 
The symbol legends are the same as those in FIGS. \ref{mu0} and \ref{VM}. The line denotes Eq. (\ref{dilatation}).}
\label{Idnu}
\end{figure}

First, we discuss the dependence of the volume fraction to the inertial number.
Surprisingly, decrease of the volume fraction caused by shear flow is also described by a power-law.
\begin{equation}
\label{dilatation}
\nu_0-\nu=s_2I^{\delta},
\end{equation}
where $\nu_0$ is the volume fraction in the $\gamma\rightarrow 0$ limit.
Note that the constants $s_2$ and $\delta$ do not depend on the details of the model.
Figure \ref{Idnu} shows that all of the data obtained in our model collapse on Eq. (\ref{dilatation}) 
with $s_2\simeq0.11$ and $\delta=0.56\pm 0.02$.
This dilatation law also illustrates ubiquity of power-law in granular materials.

The next point we wish to discuss is the relation between the present result and power-law rheology in systems at constant volume.
In particular, Xu and O'Hern \cite{xu} found a power-law relation between the shear stress and the shear rate 
in a two dimensional granular material consisting of frictionless particles.
They estimated the exponent to be $0.65$.
However, at constant volume condition, the pressure also depends on the shear rate 
so that the behavior of the friction coefficient is generally different from that of the shear stress.
Therefore, power-law friction in systems under constant volume condition are not directly related to the present result.
See Ref. \cite{hatano} for more detailed discussions on this subject.

The third point we wish to discuss is the effect of dimensionality.
In contrast to the present study, da Cruz et al. \cite{dacruz} obtained a linear friction law in a two dimensional system.
The difference may be attributed to the dimensionality of the systems, which affect the nature of contacts between particles.
In particular, the angular distribution of the tangential force is strongly anisotropic in two dimensional systems, 
while such anisotropy is not observed in our three dimensional system.
Accordingly, in the case of frictionless particles, their system exhibited a friction law that is quite similar to ours.

As the fourth point of interest, we discuss relevance of our result to earthquake mechanics 
by comparing it to a friction law recently proposed by Jop et al. \cite{jop}, 
which seems to be validated in experiments on inclined plane flow.
We stress that such flow is characterized by relatively large inertial number (typically $I \gtrsim 10^{-1}$), 
while our simulation involves much smaller $I$ values ($I\gtrsim10^{-4}$) as shown in FIGS. \ref{mu0} and \ref{VM}.
In short, Eq. (\ref{powerlaw}) involves much smaller $I$ region than Jop et al. have investigated.

Such small inertial numbers correspond to a typical configuration of seismic motion of faults.
For example, in the case that $d=1$ mm, $V=1$ m/s, $L_z=4$ cm, and $P= 100$ MPa, 
the corresponding inertial number is $10^{-4}$.
However, one may wonder that the friction law Eq. (\ref{powerlaw}) cannot lead to stick-slip motion of faults 
because the friction law found here is velocity-strengthening.
Recall that we discuss exclusively stationary-state dynamic friction.
Taking static friction into account, unstable slip is inevitable 
because static friction is always stronger than dynamic friction, 
which is mainly due to dilatation.
Therefore power-law friction in stationary states does not contradict unstable slip on faults.

The author gratefully acknowledges helpful discussions with Hisao Hayakawa, 
Namiko Mitarai, Michio Otsuki, Shin-ichi Sasa, and Masao Nakatani.


\begin{thebibliography}{99}
\bibitem{note}
The numerical data such as the friction coefficient and the volume fraction are determined as follows.
After reaching a steady state, instantaneous fluctuations of these quantities are recorded for a certain duration, 
in which the macroscopic strain of the system amounts to $40$ \%.
The average and the mean square deviation in this period are employed as the mean value and the error bar.
\end{thebibliography}
\end{document}